\documentclass[prl,twocolumn,showpacs,superscriptaddress,preprintnumbers]{revtex4}
\usepackage{amssymb}
\usepackage{graphicx}
\usepackage{dcolumn}
\usepackage{bm}
\usepackage[bookmarks = true, pdfpagemode = None, pdfstartview = FitH,colorlinks = true,
citecolor = black, linkcolor = black, urlcolor = black]{hyperref}
\usepackage{url}

\newcommand{\ket}[1]{\left\vert{#1}\right\rangle}
\newcommand{\iqp}{\left|I_q^p\right|}
\newcommand{\iqpt}{\left|I_q^p(t)\right|}
\newcommand{\iqpphi}{\left|I_q^p(\Phi_{\text{cjj}}^x)\right|}

\begin{document}

%\preprint{DWAVE/MRTNoise-10}
\title{Implementation of a Quantum Annealing Algorithm Using a Superconducting Circuit}
\author{R.~Harris}
\email{rharris@dwavesys.com}
\affiliation{D-Wave Systems Inc., 100-4401 Still Creek Dr., Burnaby, BC V5C 6G9, Canada}
\homepage{www.dwavesys.com}
\author{A.J.~Berkley}
\author{J.~Johansson}
\author{M.W.~Johnson}
\author{T.~Lanting}
\author{P.~Bunyk}
\author{E.~Tolkacheva}
\author{E.~Ladizinsky}
\affiliation{D-Wave Systems Inc., 100-4401 Still Creek Dr., Burnaby, BC V5C 6G9, Canada}
\author{B.~Bumble}
\author{A.~Fung}
\author{A.~Kaul}
\author{A.~Kleinsasser}
\affiliation{Jet Propulsion Laboratory, California Institute of Technology, Pasadena CA, USA}
\author{S.~Han}
\affiliation{Department of Physics and Astronomy, University of Kansas, Lawrence KS, USA}

\date{\today }

\begin{abstract}
A circuit consisting of a network of coupled compound Josephson junction rf-SQUID flux qubits has been used to implement an adiabatic quantum optimization algorithm.  It is shown that detailed knowledge of the magnitude of the persistent current as a function of annealing parameters is key to implementation of the algorithm on this particular type of hardware.  Experimental results contrasting two annealing protocols, one with and one without active compensation for the growth of the qubit persistent current during annealing, are presented in order to illustrate this point.
\end{abstract}

\maketitle

The successful implementation of any solid state quantum information processor will ultimately depend upon having defeated several critical challenges.  While noise \cite{noise} and fabrication variability \cite{synchronization} arguably gate progress at the moment, it must be recognized that practical issues related to scalability, architecture and algorithms also deserve attention.  It may be within experimental grasp to address some of these issues in the context of adiabatic quantum computation \cite{QA,Farhi} using state of the art designs and fabrication methods.  In this article, we address several practical details concerning the implementation of an adiabatic quantum optimization algorithm using a network of coupled rf-SQUID flux qubits \cite{architecture}.  It is shown that detailed knowledge of qubit properties as a function of annealing parameters is key to successful implementation.  Experimental results from a chain of six coupled qubits subjected to two different annealing protocols are presented in order to illustrate this latter point.

The work presented herein focuses on hardware designed to enable a particular adiabatic quantum optimization algorithm \cite{Farhi} for computing the vector $\vec{s}$ that minimizes the objective function
\begin{equation}
E(\vec{s})=-\sum_{i=1}^N h_i s_i + \sum_{i<j} K_{ij} s_i s_j \;\; ,
\label{eqn:classical}
\end{equation}

\noindent where $s_i = \pm 1$ and $N$ is the length of $\vec{s}$.  Here, $h_i$ and $K_{ij}$ are dimensionless real numbers that arise from a particular choice of problem instance.  This type of problem is of interest as it is known to be NP-hard \cite{Boros_and_Hammer}.  Equation (\ref{eqn:classical}) can be recast as the potential energy of a system of coupled spin-1/2 particles via the substitution $s_i\rightarrow\sigma_z^{(i)}$, where $\sigma_{z(x)}^{(i)}$ is the $z(x)$ Pauli matrix for spin $i$.  Let the eigenstates of $\sigma_z$ be denoted by $\ket{\uparrow}$ and $\ket{\downarrow}$.  The vector $\vec{s}$ that minimizes Eq.~(\ref{eqn:classical}) is then encoded in the groundstate $\ket{g_F}\equiv\ket{s_1\ldots s_N}$ of an Ising spin glass.  The algorithm for finding $\ket{g_F}$ relies upon exploiting the transverse ($\sigma_x^{(i)}$) degrees of freedom of a {\it quantum} Ising spin glass.  Let the Hamiltonian of such a system be
\begin{equation}
{\cal H}(t)=A(t) H_I + B(t) H_F
\label{eqn:Hs}
\end{equation}

\vspace{-0.1in}
\begin{displaymath}
H_I=-\sum_{i=1}^N \sigma_x^{(i)} \;  , \;  H_F=-\sum_{i=1}^N h_i \sigma_z^{(i)} + \sum_{i<j} K_{ij} \sigma_z^{(i)} \sigma_z^{(j)}
%\label{eqn:HiHf}
\end{displaymath}

\noindent where $0<t<t_f$, $A(0)/B(0) \gg 1$ and $A(t_f)/B(t_f) \ll 1$.  We refer to $A(t)$ and $B(t)$, which have units of energy, as {\em envelope functions}.  If $A(0)$ is much larger than all other relevant energy scales, including temperature, then the system will begin the evolution at $t=0$ in the groundstate of $H_I$, $\ket{g_I}\equiv\prod_{i=1}^N\left(\ket{\downarrow}+\ket{\uparrow}\right)/\sqrt{2}$, with a probability of 1.  If the subsequent evolution is adiabatic, then the system will be found in $\ket{g_F}$ at $t=t_f$.  Note that the $t$-dependence of this algorithm is entirely contained in $A(t)$ and $B(t)$, which are explicitly {\it not} functions of $N$ or problem instance.

The most convenient forms for the envelope functions depend upon the details of the hardware.  Consider a network of coupled compound Josephson junction (CJJ) rf-SQUID qubits \cite{CJJ,synchronization} whose persistent currents $\iqpphi$ and tunneling energies $\Delta(\Phi_{\text{cjj}}^x)$ are identical as a function of CJJ bias $\Phi_{\text{cjj}}^x$.  If $\Phi_{\text{cjj}}^x$ is a function of $t$, then the processor Hamiltonian can be expressed as 
\begin{equation}
\label{eqn:Hprocessor}
{\cal H}(t) = -\sum^N_{i=1}\frac{1}{2}\left[\epsilon_i(t)\sigma_z^{(i)}+\Delta(t)\sigma_x^{(i)}\right]+\sum_{i<j}J_{ij}(t)\sigma_z^{(i)}\sigma_z^{(j)}
\end{equation}

\noindent where $\epsilon_i(t)\equiv2\iqpt\Phi_i^x(t)$ represents the energy bias of a flux qubit ($-\Phi_0\leq \Phi_{\text{cjj}}^x(t)\leq -\Phi_0/2$, $\Phi_0\equiv h/2e$), $\Phi_i^x(t)$ is an externally controlled flux bias and $J_{ij}\equiv M_{ij}\iqpt^2$ represents the pairwise coupling mediated by a mutual inductance $M_{ij}$.  Comparison of Eqns.~(\ref{eqn:Hs}) and (\ref{eqn:Hprocessor}) readily yields one envelope function: $A(t)=\Delta(t)/2$.  On the other hand, there is no unique definition of $B(t)$.  One approach is to scale $\epsilon_i(t)$ and $J_{ij}(t)$ by a convenient factor: let $B(t)=J_{\text{AFM}}(t)\equiv M_{\text{AFM}}\iqpt^2$, where $M_{\text{AFM}}$ is the strongest antiferromagnetic (AFM) coupling needed to embed a particular problem instance.  Doing so implies
\begin{equation}
\label{eqn:HfMap}
H_F=-\sum_{i=1}^N\frac{\Phi_i^x(t)}{M_{\text{AFM}}\iqpt}\sigma_z^{(i)}+\sum_{i<j}\frac{M_{ij}}{M_{\text{AFM}}}\sigma_z^{(i)}\sigma_z^{(j)} \; ,
\end{equation}

\noindent which yields a prescription for mapping Eq.~(\ref{eqn:classical}) onto the hardware: $h_i=\Phi_i^x(t)/M_{\text{AFM}}\iqpt$ and $K_{ij}=M_{ij}/M_{\text{AFM}}$.  Note that $K_{ij}$ has no $t$-dependence.  To avoid $t$-dependence in $h_i$ one must use a $t$-dependent qubit flux bias:
\begin{equation}
\label{eqn:hdef}
\Phi_i^x(t)\equiv h_iM_{\text{AFM}}\iqpt \;\; .
\end{equation}

\noindent  We denote annealing processes that use $\Phi_i^x(t)$ as defined by Eq.~(\ref{eqn:hdef}) as {\it controlled annealing}.  Processes in which $\Phi_i^x$ are held static will be termed {\it fixed annealing}.  Formally, the relative contributions of $\sigma_z^{(i)}$- versus $\sigma_z^{(i)}\sigma_z^{(j)}$-terms to Eq.~(\ref{eqn:Hprocessor}) will vary during fixed annealing.  It will be demonstrated that fixed annealing is not a viable means of implementing an optimization algorithm and that controlled annealing remedies the problem cited above.

\begin{figure}[tbp]
\includegraphics[width=2.75in]{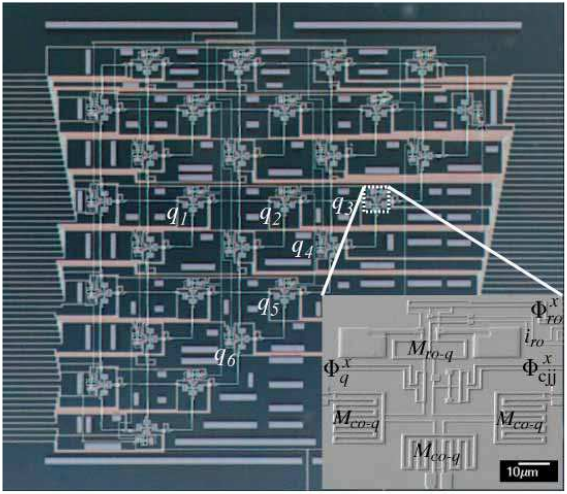}
\caption{(Color online)  Optical image of circuit with qubits $q_1\ldots q_6$ labeled.  Inset shows an electron micrograph of $q_3$ with bias lines for controlling qubit CJJ bias ($\Phi_{\text{cjj}}^x$), qubit flux bias ( $\Phi_q^x$), readout flux bias ($\Phi_{\text{ro}}^x$) and readout current bias ($i_{\text{ro}}$) as indicated.  Transformers between qubit and couplers denoted as $M_{\text{co}-q}$.  Transformer between qubit and readout denoted as $M_{\text{ro}-q}$.}
\label{fig:Leda}
\end{figure}

\begin{figure}[tbp]
\includegraphics[width=2.8in]{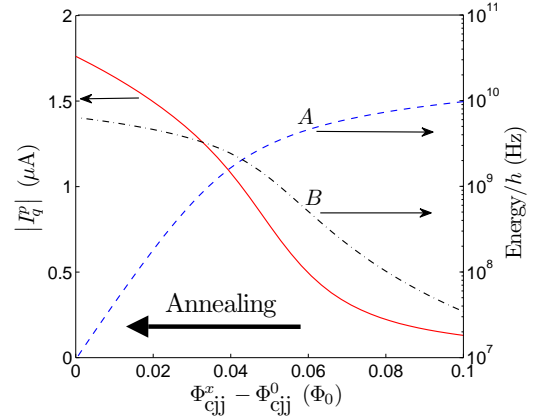}
\caption{(Color online)  CJJ bias dependence of $\iqp$ and envelope functions $A=\Delta/2$ and $B=J_{\text{AFM}}\equiv M_{\text{AFM}}\iqp^2$.  Direction of CJJ bias ramp during annealing as indicated.}
\label{fig:qubitparameters}
\end{figure}

We have performed a series of annealing experiments on a chain of 6 CJJ rf-SQUID qubits with intervening AFM couplers.  These particular qubits, as depicted in Fig.~\ref{fig:Leda} and labeled as $q_1\ldots q_6$, were part of a larger circuit and were selected on account of their low Josephson junction asymmetry \cite{synchronization}.  The chip was fabricated on an oxidized Si wafer with Nb/Al/Al$_2$O$_3$/Nb trilayer junctions and three Nb wiring layers separated by sputtered SiO$_{2}$.  It was mounted to the mixing chamber of a dilution refrigerator and cooled to $T=35\,$mK in the presence of a very low ($<9\,$nT) background magnetic field inside a PbSn coated shield.  Each qubit was connected to three others via in-situ tunable rf-SQUID couplers, which we treat as classical effective mutual inductances \cite{coupler}.  Each qubit was also inductively coupled to a hysteretic dc-SQUID for readout \cite{readout} The chain of qubits studied herein was isolated from the rest of the chip by tuning unused couplers to provide zero coupling and biasing unused qubits with $\Phi_{\text{cjj}}^x=-\Phi_0/2$ to minimize their persistent currents.  The CJJ bias dependences of $\iqp$ and $\Delta$ of these qubits were synchronized by applying the methods described in Ref.~\cite{synchronization}.  Further details regarding qubit calibration and measurements of $\iqpphi$ and $\Delta(\Phi_{\text{cjj}}^x)$ can be found therein.  The expected CJJ bias dependence of the envelope functions, as determined from the mean device parameters reported in Ref.~\cite{synchronization}, has been plotted in Fig.~\ref{fig:qubitparameters}.  Here, the synchronization CJJ bias $\Phi_{\text{cjj}}^0$ has been defined such that $\Delta(\Phi_{\text{cjj}}^0)/2h=10\,$MHz. 

\begin{figure}[bp]
\includegraphics[width=2.8in]{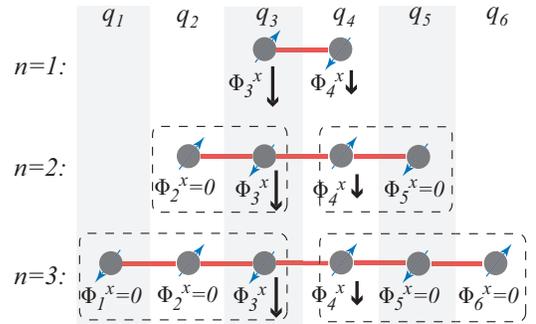}
\caption{(Color online)  AFM coupled domain experiments.  Qubits represented as spin-1/2 particles (circles with arrows) and AFM couplers as solid lines.  $n$ denotes number of qubits per AFM domain (dashed boxes).  Flux biases $\Phi_i^x$ as noted.}
\label{fig:Domains}
\end{figure}

To study controlled versus fixed annealing, we have measured three different configurations of the chain, as shown schematically in Fig.~\ref{fig:Domains}.  The diagram depicts a sequence of experiments in which qubits $q_3$ and $q_4$ were subjected to variable flux biases while all other qubit flux biases $\Phi_i^x=0$.  Qubits that were not used in a particular experiment had their CJJ biases held at $\Phi_{\text{cjj}}^x=-\Phi_0/2$ so as to decouple them from the active qubits.  In each successive experiment, one more AFM coupled qubit was activated on both ends of the chain.  In the limit $\Delta\ll J_{\text{AFM}}$, one can write analytical solutions for the eigenstates of Eq.~(\ref{eqn:Hprocessor}) for each of the experiments depicted in Fig.~\ref{fig:Domains}.  The results show that the four lowest energy levels can be ascribed to a system of two AFM coupled {\it effective} qubits with the same $\iqp$ as a single qubit, but a renormalized tunneling energy $\Delta_{\text{eff}}\approx \Delta^n/(2J_{\text{AFM}})^{n-1}$, where $n$ is the number of qubits in an AFM domain:
\begin{eqnarray}
\label{eqn:Heff}
{\cal H}(n,t) & \approx & -\sum_{i=3}^4\frac{1}{2}\left[\epsilon_i(t)\sigma_z^{(i)}
+\Delta_{\text{eff}}(n,t)\sigma_x^{(i)}\right] \nonumber\\ 
 & & +J_{AFM}(t)\sigma_z^{(3)}\sigma_z^{(4)}
\end{eqnarray}

\noindent Thus, the experiments depicted in Fig.~\ref{fig:Domains} were isomorphic 2-qubit optimization problems in which the tunneling energy $\Delta_{\text{eff}}$ was a strong function of $n$.  An optimization algorithm applied to these systems ought to yield results (states of $q_3$ and $q_4$) that are independent of $n$.

\begin{figure}[tbp]
\includegraphics[width=3.2in]{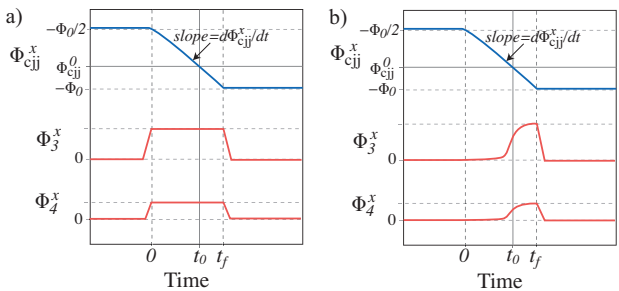}
\caption{(Color online)  Waveforms for a) fixed and b) controlled annealing.  All active qubits utilize the same $\Phi_{\text{cjj}}^x$ waveform.  Only qubits $q_3$ and $q_4$ carried $t$-dependent $\Phi_i^x$.}
\label{fig:waveforms}
\end{figure}

The fixed and controlled annealing waveform patterns are depicted in Fig.~\ref{fig:waveforms}a and \ref{fig:waveforms}b, respectively.  In both cases, the CJJ biases of all active qubits were simultaneously ramped from $-\Phi_0/2$ to $-\Phi_0$.  The ramps were digitally low pass filtered using $f_c=1.3\,$MHz so as to avoid uncontrolled delays between qubits due to the limited bandpass of the wiring \cite{synchronization}.  The slope of the ramps in the vicinity of $\Phi_{\text{cjj}}^x(t_0)=\Phi_{\text{cjj}}^0$ was $d\Phi_{\text{cjj}}^x/dt=3.2\times10^{-2}\,\Phi_0/\mu$s.  For fixed annealing, $\Phi_3^x$ and $\Phi_4^x$ were set to static values during the CJJ ramps.  For controlled annealing, $\Phi_3^x(t)$ and $\Phi_4^x(t)$ carried scaled time dependent waveforms as dictated by Eq.~(\ref{eqn:hdef}).  Here, $\iqpphi$ was obtained by sampling the smooth function shown in Fig.~\ref{fig:qubitparameters}.  Choosing the target values of $h_3=\epsilon_3/2J_{\text{AFM}}$ and $h_4=\epsilon_4/2J_{\text{AFM}}$ and knowing $M_{\text{AFM}}=1.35\pm0.02\,$pH from independent measurements \cite{synchronization} allowed for construction of $\Phi_3^x(t)$ and $\Phi_4^x(t)$.  In both fixed and controlled annealing,  these fluxes were set to zero during qubit initialization and at the end of annealing prior to readout.

\begin{figure}[tbp]
\includegraphics[width=2.0in,bb=100 220 490 590]{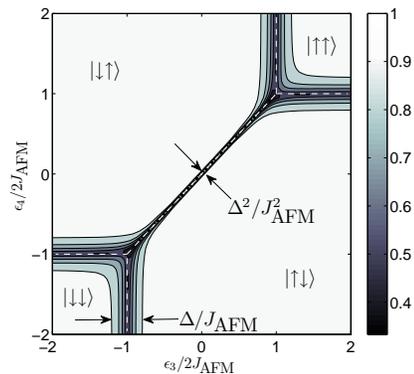}
\caption{(Color online)  Most probable state map $\left|\left<g|q_3q_4\right>\right|^2$ for the $n=1$ experiment in the limit $\Delta\ll J_{\text{AFM}}$.  Dashed lines indicate phase boundaries in the limit $\Delta\rightarrow 0$.}
\label{fig:stability}
\end{figure}

To develop a sense of how $\Delta_{\text{eff}}(n)$ impacts the dynamics, consider the $n=1$ experiment as a function of $\epsilon_3/2J_{\text{AFM}}$ and $\epsilon_4/2J_{\text{AFM}}$ in the limit $\Delta\ll J_{\text{AFM}}$.  The results of a calculation of the probability $\left|\left<g|q_3q_4\right>\right|^2$, where $\ket{g}$ is the groundstate of Eq.~(\ref{eqn:Heff}), are shown in Fig.~\ref{fig:stability}.  Here, the AFM states $\ket{\downarrow\uparrow}$ and $\ket{\uparrow\downarrow}$ occupy the majority of the plane and intersect at a diagonal phase boundary.  Horizontal and vertical phase boundaries are located at $\pm 1$ where the AFM regions intersect the ferromagnetic (FM) states $\ket{\downarrow\downarrow}$ and $\ket{\uparrow\uparrow}$.  The width of the region about the phase boundaries where one may observe quantum tunneling between states are as noted in the diagram.

Let the system depicted in Fig.~\ref{fig:stability} be subjected to fixed annealing.  In this case, $\epsilon_i(t)/2J_{\text{AFM}}(t)\propto 1/\iqpt$.  Since $\iqp$ is a monotonically increasing function of $t$, then $(\Phi_3^x,\Phi_4^x)$ corresponds to a point that moves radially inward in the $(\epsilon_3/2J_{\text{AFM}},\epsilon_4/2J_{\text{AFM}})$-plane as annealing progresses.  If this point traverses a phase boundary at a time $t_p$, then it will be forced through a phase transition.  The probability of faithfully tracking the groundstate will be a function of $\Delta(t_p)$ and of the time spent in the vicinity of the phase boundary, which will be proportional to $d\Phi_{\text{cjj}}^x/dt|_{t=t_p}$.  In this regard, fixed annealing could be a generalized Landau-Zener experiment \cite{LZ}.  Now consider replacing the single qubits $q_3$ and $q_4$ with AFM domains of size $n>1$: $\epsilon_i(t)/2J_{\text{AFM}}(t)$ will remain unchanged, but $\Delta(t)\rightarrow\Delta_{\text{eff}}(t)$.  Since the tunneling energy will be suppressed, the probability of achieving the groundstate will decrease with $n$ when all other experimental parameters are equal.  Thus, the locations of {\it apparent} phase boundaries in the $(\Phi_3^x,\Phi_4^x)$-plane will depend upon $n$.

In contrast to the above scenario, a system subjected to controlled annealing never traverses a phase boundary in the limit $\Delta\ll J_{\text{AFM}}$.  This is assured because $\epsilon_i/2J_{\text{AFM}}$ will be independent of $t$ by construction.
As such, the locations of phase boundaries in the controlled annealing $(\epsilon_3/2J_{\text{AFM}},\epsilon_4/2J_{\text{AFM}})$-plane will be {\it independent} of $n$.  

Experimental maps of the most probable final spin configuration for fixed and controlled annealing were generated by sampling on a grid of points in the $(\Phi_3^x,\Phi_4^x)$- and $(\epsilon_3/2J_{\text{AFM}},\epsilon_4/2J_{\text{AFM}})$-plane, respectively. 
Since the state of each qubit could be read by its own dedicated dc-SQUID magnetometer, it was possible to unambiguously identify the final spin configuration of any given AFM domain at the end of every annealing cycle.  Running 64 cycles per point provided histograms of the final spin configuration from which one could readily identify the most probable state.  We note that there were no measurements that ever indicated single qubit flips had occurred in any of the AFM domains.

\begin{figure}[tbp]
\begin{tabular}{cc}
\includegraphics[width=1.7in]{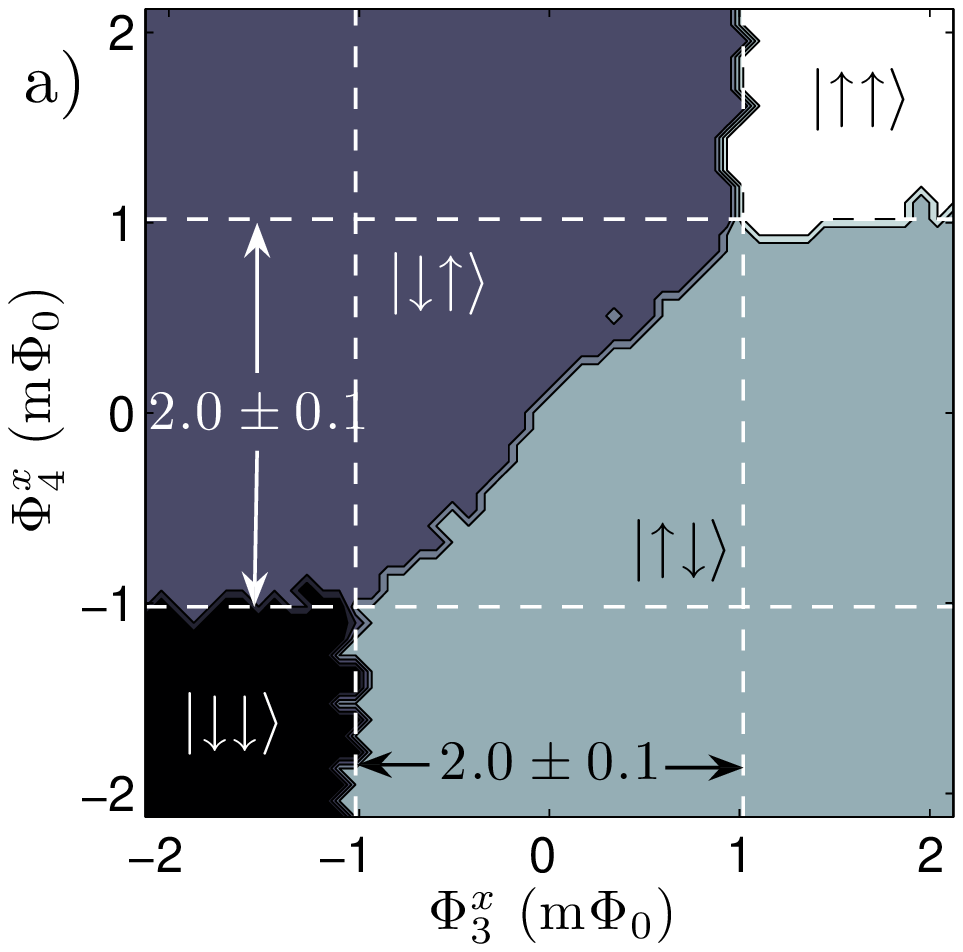} &
\includegraphics[width=1.7in]{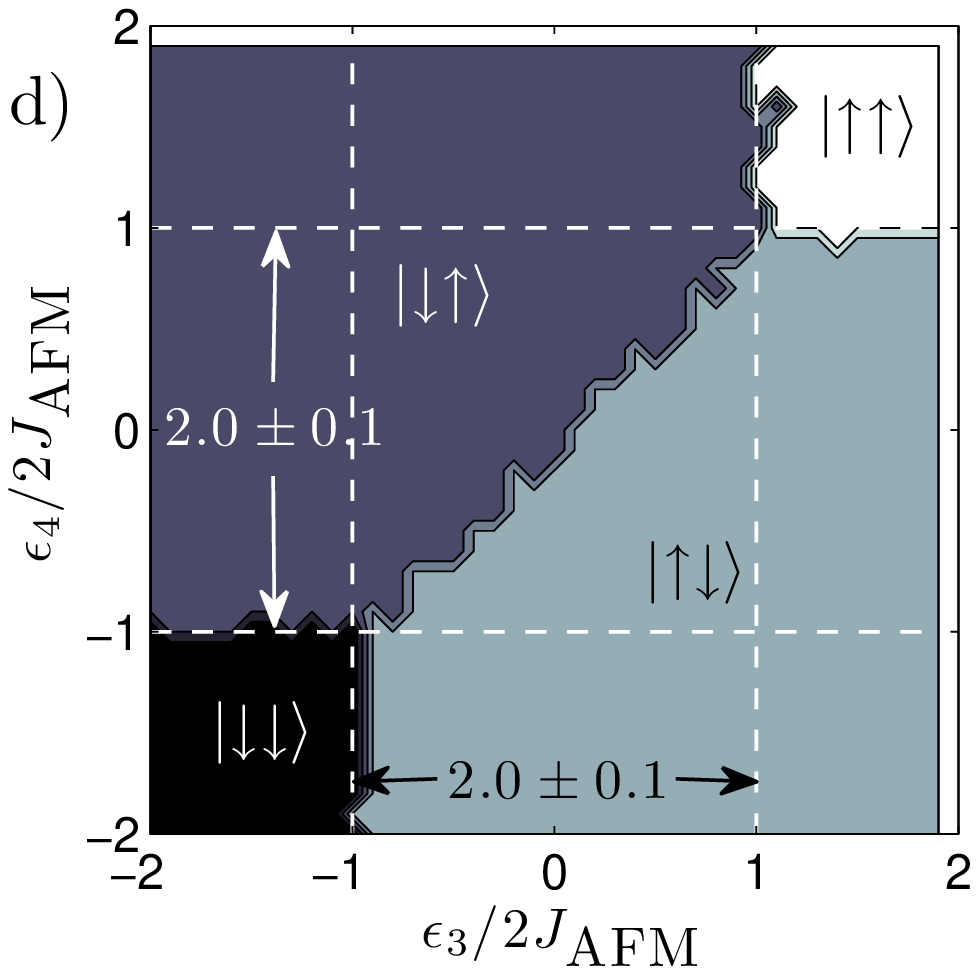} \\
\includegraphics[width=1.7in]{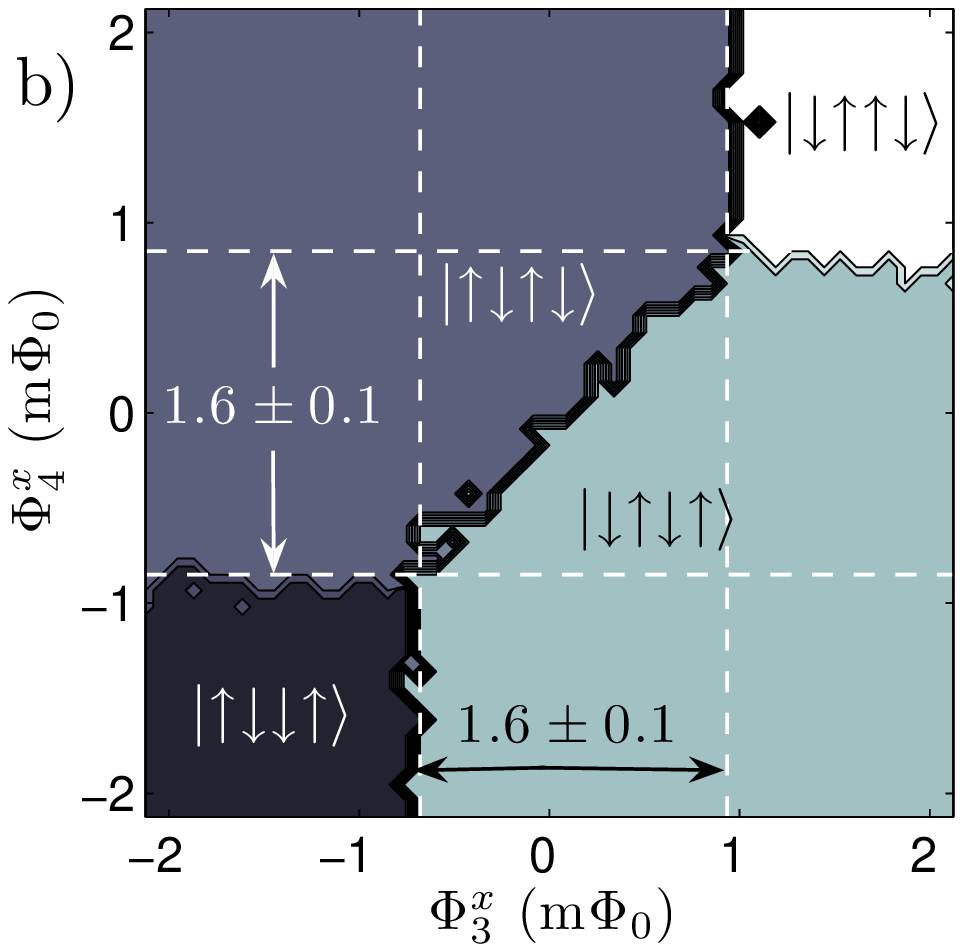} &
\includegraphics[width=1.7in]{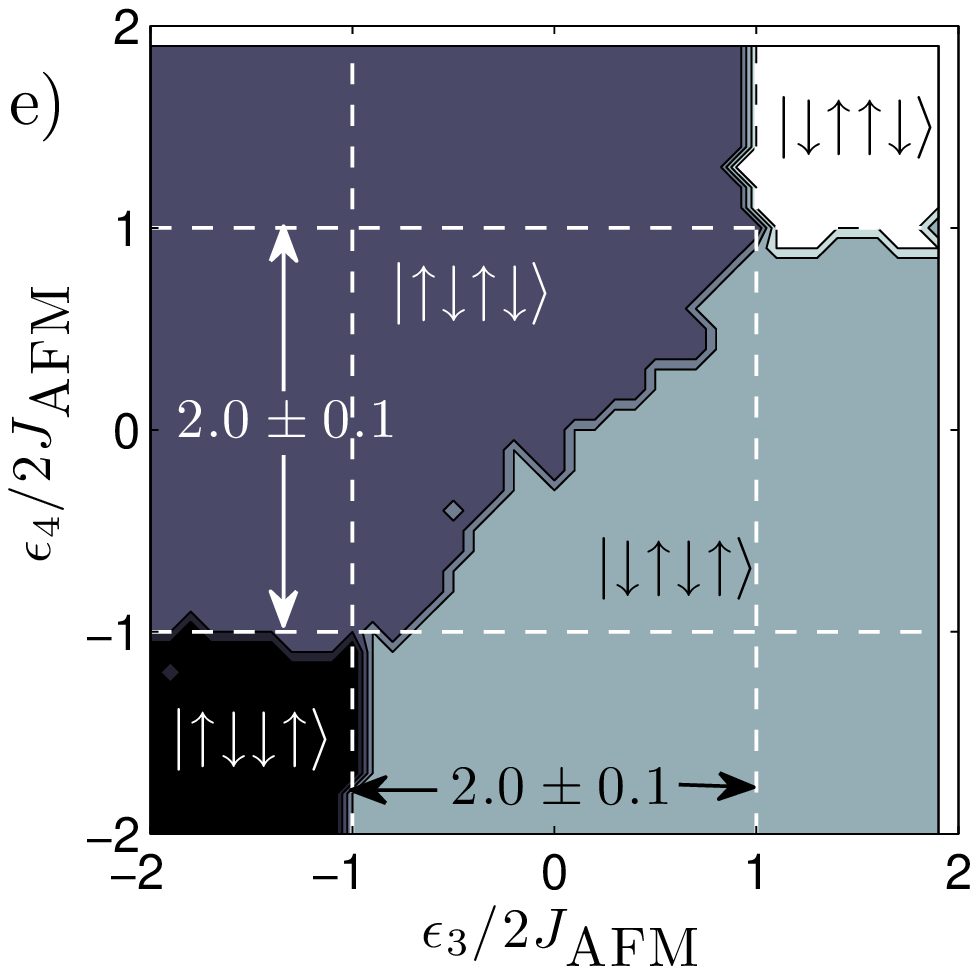} \\
\includegraphics[width=1.7in]{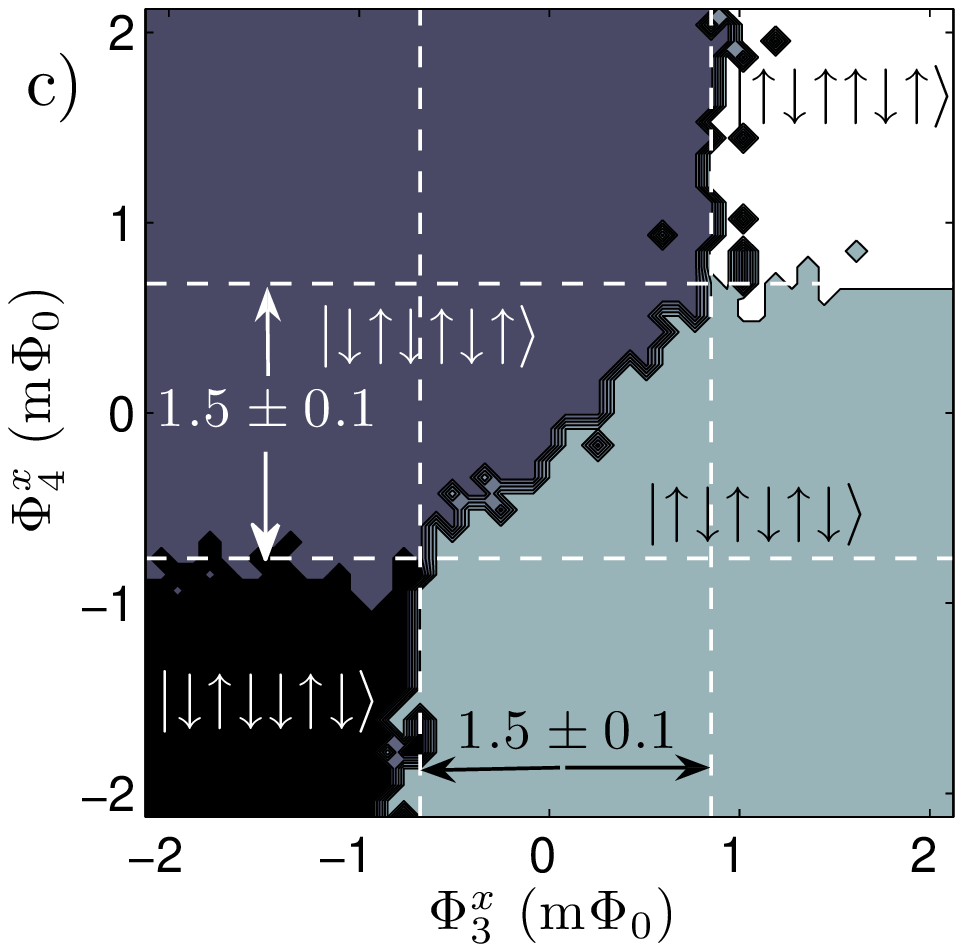} &
\includegraphics[width=1.7in]{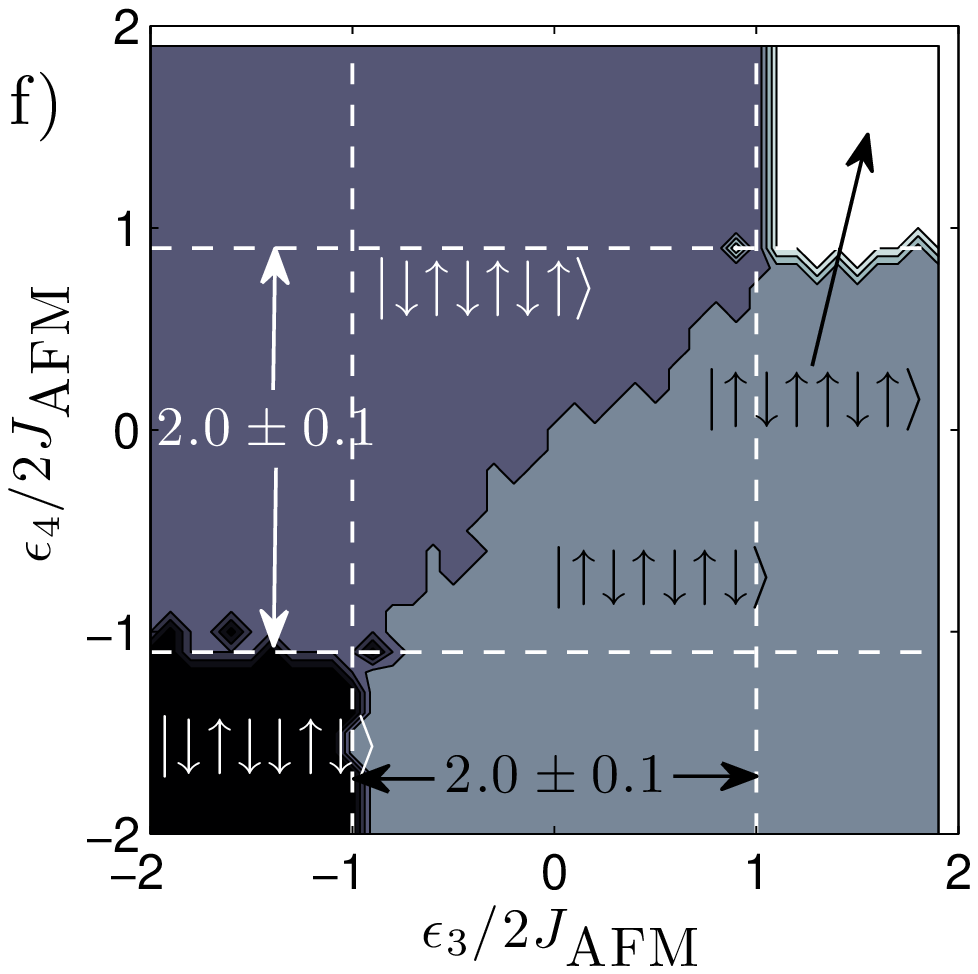}  \\
\end{tabular}
\caption{(Color online) Maps of the most probable final spin configuration as a function of fixed annealing flux biases (a, b and c) and controlled annealing dimensionless parameters (d, e and f).  States are labeled as $\ket{q_3q_4}$ (a and d), $\ket{q_2q_3q_4q_5}$ (b and e) and $\ket{q_1q_2q_3q_4q_5q_6}$ (c and f).}
\label{fig:maps}
\end{figure}

Results of the fixed annealing experiments are shown in Figs.~\ref{fig:maps}a$\rightarrow$c.  Here, we have colored each point in the $(\Phi_3^x,\Phi_4^x)$-plane to represent the spin configuration $\ket{\ldots q_3q_4\ldots}$ that was observed with the highest probability.  The data indicate that the {\it apparent} AFM interaction between $q_3$ and $q_4$ becomes weaker with increasing $n$, as evidenced by the progressive encroachment of the FM states.  The difference in flux between the pairs horizontal and vertical boundaries can be interpreted as an effective AFM coupled flux $2\Phi_q^*\equiv 4M_{\text{AFM}}\left|I_q^p(\Phi_{\text{cjj}}^*)\right|$, where $\Phi_{\text{cjj}}^*$ represents a particular CJJ bias that depends upon $n$.  Decreasing $\Phi_q^*$ indicates that the dynamics of the AFM chain effectively freeze out at larger $\Phi_{\text{cjj}}^x$ where $\iqp$ is smaller (see Fig.~\ref{fig:qubitparameters}).  A detailed study of the dependence of $\Phi_q^*$ upon $n$ and $d\Phi_{\text{cjj}}^x/dt$ may provide information regarding the mechanism by which this circuit achieves its final configuration.  This matter will be the topic of a future publication.  Nonetheless, these data clearly show that fixed annealing is not a viable means of operating this hardware as a  processor as the solution to a given isomorphic 2-qubit problem depended upon $n$.

The controlled annealing results presented in Figs.~\ref{fig:maps}d$\rightarrow$f show no dependence upon $n$.  To within experimental error, the phase boundaries in the $(\epsilon_3/2J_{\text{AFM}},\epsilon_4/2J_{\text{AFM}})$-plane agree with the results of Eq.~\ref{eqn:Heff} in the limit $g\rightarrow 0$.  Additional measurements also revealed no dependence upon $d\Phi_{\text{cjj}}^x/dt$ over 3 orders of magnitude ($3.2\times10^{-4}$ to $0.32\,\Phi_0/\mu$s).  Therefore, it has been demonstrated that the controlled annealing protocol returns the correct solutions to optimization problems posed as a set of $h_i=\epsilon_i/2J_{\text{AFM}}$ and $K_{ij}=1$, per Eq.~(\ref{eqn:classical}).  Consequently, this circuit, when used in conjunction with controlled annealing, can be viewed as a prototype quantum Ising spin glass computer.

{\it Conclusions:}  A method for annealing coupled CJJ rf-SQUID flux qubits that accounts for realistic device behavior and embodies the physics of a particular quantum adiabatic optimization algorithm has been experimentally demonstrated.  This work represents a critical step toward the development of practical adiabatic quantum information processors.

We thank J.~Hilton, G.~Rose, P.~Spear, A.~Tcaciuc, F.~Cioata, E.~Chapple, C.~Rich, C.~Enderud, B.~Wilson, M.~Thom, S.~Uchaikin, M.~Amin, F.~Brito and D.~Averin.  Samples were fabricated by the Microelectronics Laboratory of the Jet Propulsion Laboratory, operated by the California Institute of Technology under a contract with NASA.  S.Han was supported in part by NSF Grant No. DMR-0325551.

\end{document}